\documentclass[conference]{IEEEtran}
\usepackage{graphicx} 
\ifCLASSINFOpdf
\else
\fi
\usepackage{enumitem}
\usepackage{url} 
\begin{document}
%


\title{``So what if I used GenAI?" - Implications of Using Cloud-based GenAI in Software Engineering Research}
\author{\IEEEauthorblockN{Gouri Ginde}
\IEEEauthorblockA{Department of Electrical and Software Engineering\\
University of Calgary\\
Email: gouri.ginde@ucalgary.ca}}


%


\maketitle

\begin{abstract}
Generative Artificial Intelligence (GenAI) advances have led
to new technologies capable of generating high-quality code, natural language, and images. The next step is to integrate GenAI technology into various aspects while conducting research or other related areas, a task typically conducted by researchers. Such research outcomes always come with a certain risk of liability. This paper sheds light on
the various research aspects in which GenAI is used, thus raising awareness of its legal implications to novice and budding researchers. In particular, there are two risks: data protection and copyright. Both aspects are crucial for GenAI. We summarize key aspects regarding our current knowledge that every software researcher involved in using GenAI should be aware of to avoid critical mistakes that may expose them to liability claims and propose a checklist to guide such awareness. 
\end{abstract}


%
\IEEEpeerreviewmaketitle

\begin{figure*}[!h]
    \centering
\includegraphics[scale=.45]{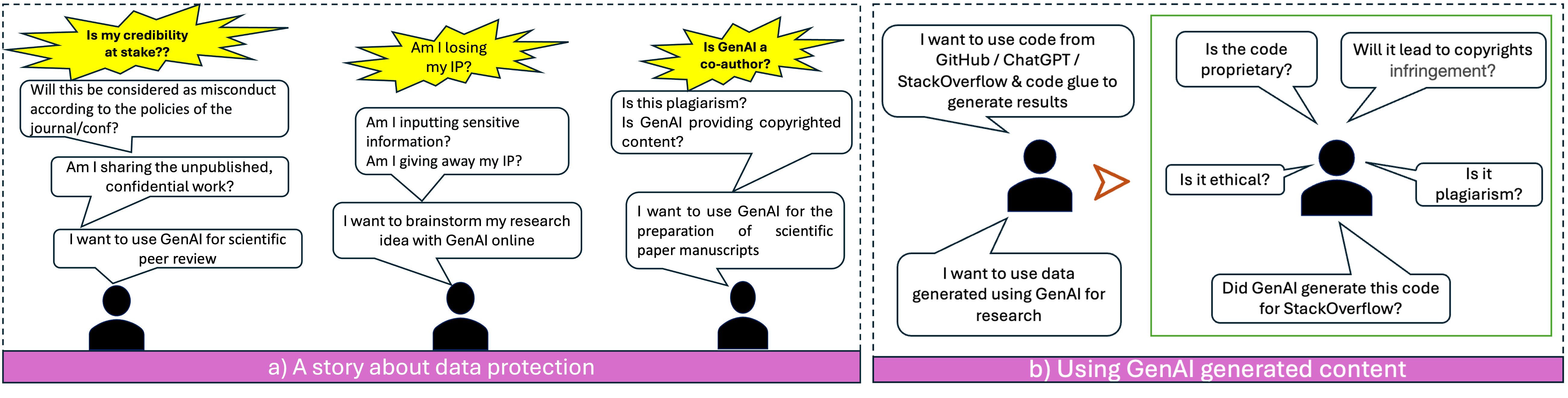}
\vspace{-5mm}
    \caption{Conceptual scenarios when legal implications of GenAI should be considered while doing research}
    \label{scenario}
\end{figure*}
\section{Introduction}
Generative AI (GenAI) is revolutionizing software development and software engineering (SE) in gereral more profoundly than any other recent technology \cite{ebert2024hints}. It has proliferated into research-oriented aspects as well \cite{ebert2023generative}. At the core of GAI are large language models (LLMs), vast neural networks trained on massive text datasets. Approximately 60\% of GenAI applications are utilized during the software development phase \cite{ebert2024hints}. However, it is important to note that any input provided—whether a simple prompt or content-specific details—can contribute to the ongoing evolution of these typically cloud-based LLMs. Conversely, the outputs generated by LLMs are not guaranteed to be free from intellectual property rights (IPR) considerations, posing potential legal and ethical challenges. Thus, while using cloud-based GenAI services, software researchers must be careful of its implications on their research outcomes. 

GenAI can be used in various aspects of SE research such as  scholarly paper reviewing, brainstorming and ideating using GenAI, writing manuscripts, using data generated using LLMs, and programming and developing (source) code. A few of the dilemmas faced by the researchers are shown in Figure \ref{scenario}. The ownership of the content generated by the model and the use of third-party content within the generated elements (reviews/code/manuscript) are also significant ethical concerns. For example, while GenAI can assist editors and peer reviewers in completing repetitive or tedious tasks, there is a risk that it may not mitigate existing biases and that human judgment calls are still necessary. 

The academic community has been rattled by the free access to GenAI services usage in research. Well aclaimed conference ICML wrote the following as part of thier policies:  ``The Large Language Model (LLM) policy for ICML 2023 prohibits text produced entirely by LLMs (i.e., “generated”). Similarly, ICSE and other conferences in SE have similar guidelines for LLM usage policy. Although this does not prohibit authors from using LLMs for editing or polishing author-written text, it is evident that the GenAI tools raise concerns about research transparency, reproducibility, and potential biases that threaten scientific integrity and equity in research outcomes. 

\textbf{Our objective in this study is to understand the risks and implications of using cloud-based GenAI in SE research and propose a checklist to overcome or mitigate the possible risks associated with GenAI's utilization in research.}

GenAI is a broader category of AI tools designed to generate new content such as images, text, video, code and audio. In this paper, we will focus on LLMs, which are specifically designed to process and generate human language. LLMs can be grouped into three types: open-sourced, enterprise and free-tier access \cite{OpenSour82:online}, \cite{OpenSour74:online}, \cite{alipour2024chatgpt}. Table \ref{comparison} provides detailed information on these types and their differences.

\begin{table*}[!h]
\vspace{-2mm}
\renewcommand{\arraystretch}{1.2}
\caption{ Summary of comparison of various types of LLMs  \cite{OpenSour82:online}, \cite{OpenSour74:online}, \cite{alipour2024chatgpt}}
\begin{tabular}{p{1.6cm}|p{5.25cm}|p{5.25cm}|p{5.25cm}}
\multicolumn{1}{c}{\textbf{}} & \multicolumn{1}{c}{\textbf{Open-Source}}    & \multicolumn{1}{c}{\textbf{Free-Tier (cloud-based)}} & \multicolumn{1}{c}{\textbf{Enterprise (license-based)}}       \\
\textbf{Accessibility}         & Publicly available with open licensing           & Limited free access via APIs/interfaces         & Subscription-based access            \\
\textbf{Customization}         & Fully customizable; fine-tuning supported.        & Limited to prompt engineering; no fine-tuning.      & Domain-specific tuning.      \\
\textbf{Data Privacy}          & Full control; user responsibility for security   & Data may be retained for model improvements        & Guaranteed privacy \\
\textbf{Cost}                  & Free to use; hosting and training incur costs   & Free but with usage limits  & Subscription pricing          \\
\textbf{Ease of Use}           & Requires ML expertise to deploy and manage       & Beginner-friendly; no setup needed                 & Seamless integration options.    \\
\textbf{Performance}           & Varies by model; requires optimization. & Optimized for general-purpose tasks.                & High performance, tailored for business needs.      \\
\textbf{Examples}              & LLaMA, Falcon, GPT-J, GPT-NeoX, Mistral.          & ChatGPT Free, Google Bard, Claude Instant.          & ChatGPT Enterprise, Azure OpenAI
\end{tabular}
\label{comparison}
\end{table*}

\textbf{Motivation:}
Our high-level analysis on the academic and law communities on stack exchange\footnote{https://stackexchange.com/about - a network of question-and-answer (Q\&A) websites on topics in diverse fields, each community site covering a specific topic, where questions, answers, and users are subject to a reputation award process. Comprises over 173 Q\&A communities}, a widely used question-answering platform in SE \cite{wang2018understanding}, showed some intriguing findings where mostly free-tier (cloud-based) GenAI tools are discussed.
Of the 45,000 questions posted on the academic stack exchange, over 2,000 are on ethics, 960 on plagiarism, and 180 on research misconduct. However, only 45 questions with the Generative-AI tag are posted. The law stack exchange with an artificial intelligence tag had 81 questions. A closer look at these questions (a few listed below) and their views (V) raise serious concerns about research's legal and ethical aspects.

\begin{itemize}[leftmargin=*]
    \item Should I report a review I suspect to be AI-generated? (5k views)
    \item What should I do if I suspect one of the journal reviews I got is AI-generated? (27K views)
    \item Co-author uses ChatGPT for academic writing - is it ethical? (18K views)
    \item Should I preemptively confess after submitting work that was partially generated by ChatGPT? (9k views)
    \item Are there examples of journals with an explicit policy on GPT-3 and equivalent language models? (2K views)
    \item Is it OK to generate parts of a research paper using a large language model such as ChatGPT? (23K views)
    \item Is Utilizing AI Tools for Conducting Literature Reviews in Academic Research Advisable? (9K views)
    \item How can a programming beginner effectively utilize ChatGPT as a tool for programming? (338 views)
    \item Is there any automated system available that validate the accuracy of the data generated by GenAI? (397 views)
    \item Is the generated code by Code-Llama and LLama-2 models licensed somehow or it has copyright issues? (287 views)
\item Copyright risks for code contributed by generative AI (699 views)
\item Do Llama-2 and Code-Llama models collects my code? (422 views)

\end{itemize}
These questions show evidence for a widespread curiosity regarding the use of GenAI in academic writing, reviewing, and research processes in general; however, there is little interest regarding the copyright, licensing, and ethical issues about GenAI's utilization in research (last four questions) although its repercussions can be detrimental. This motivated us to explore the ethical implications of using GenAI in research, especially in the SE, as the data, code, and evaluation are the core of it. 

\section{What are the risks and implications of using GenAI in SE research?}
Recently, legal implications of using GenAI have been explored in domains such as media \cite{bayer2024legal} and medical education \cite{marz2024legal} \cite{khlaif2023potential}. GenAI has exponentially triggered unethical news, gossip outlets, and disinformation networks in the media domain, making it impossible for the general public to differentiate weed from chaff. The medical education study emphasizes that the rapid development, adoption and use of AI technologies in healthcare requires healthcare professionals to master experimental techniques, even if they are not yet recognized as standards. 
Similar implications and risks exists for SE domain and we list them as follows.\\
\textbf{Data Privacy and Security:} Researchers have been increasingly using GenAI such as ChatGPT to assist in ideation \cite{gozalo2023survey} due to its ability to act as a conversatioal agent. LLM models such as GALACTICA \cite{taylor2022galactica}  and MINERVA \cite{lewkowycz2022solving} can store, combine and reason about scientific language have shown remarkable results. However, researchers might overlook the terms of services (TOS) which clearly mention that the unless opted out, they may use content to train their future models to provide, maintain, develop, and improve their services \cite{Termsofu98:online}. Thus, one might give away their ideas and highly private and sensitive information to a unknown entity . Additionally, in a recent study conducted at 2024 Neural Information Processing Systems (NeurIPS) conference on the use of LLMs in the scientific peer review process aiding scientific peer review \cite{goldberg2024usefulness} demonstrated that the over 70\% authors found it useful and were willing to revise their papers based on the feedback. However, it was cautioned that this approach has serious data privacy and security concerns. \\
While there is a debate regarding if it acceptable to using an AI to do the bulk of writing Latex \cite{mathemat88:online} as it would be considered cheating in the same way that the use of a calculator would be looked upon as cheating if you were being tested on your ability with mental arithmetic. However, the data privacy and security are still at stake as the work is still unpublished. \\
\textbf{Licensing Issues:} Noticing the gravity of the issue, Stackoverflow (SO) made a strict policy against use of GenAI use while posting answers \cite{PolicyGe94:online} as they observed that there are SO users active for years that previously produced only few answers now posting over 50 in less than a day. The amount of AI generated answers could suffocate SO if everyone starts doing it without giving proper credit to the AI. Contributors tend to present the content as their own, thus misrepresenting someone else's work.
GenAI models are trained on massive scale datasets available online. However, not all reveal the source of their training data which has raised wide scaled upraor against copyright violations. Recent innovation of GitHub copilot came under hammer due to such as allegation \cite{GitHubCo41:online} which clearly explains that the Training AI systems on public GitHub repositories, and potentially additional sources, has led to the violation of legal rights of many creators who posted code or other work under specific open-source licenses on GitHub. These licenses include 11 popular open-source licenses, such as the MIT, GPL, and Apache, all requiring proper attribution of the author's name and copyright. \\
 \textbf{Academic integrity: }It's debatable whether LLMs like ChatGPT are suitable for editing and polishing text. For example, interviewed by the Verge \cite{ChatGPTa64:online}, Deb Raji (AI research fellow, Mozilla Foundation) highlightes that LLM) differ from tools like Grammarly, as they are not solely designed for text refinement but also generate novel content, including potentially problematic outputs like spam. This makes them more complex and distinct from simpler corrective tools. Similarly, COPE (Committee on Publication Ethics) \cite{AboutCOP44:online} is a international body which is committed to educating and supporting editors, publishers, universities, research institutes, and all those involved in publication ethics warns that critics have recently suggested that using AI for review purposes risks putting confidential information back into the public domain. \\
\textbf{Copyright and Intellectual Property:} SO has put the 2-year-old ban on using GenAI for coding in SO answers, however, they have been struggling to identify and put heavy moderating efforts towards it \cite{PolicyGe94:online}. Moderator's post regarding this which is viewed  1.4m times emphasizes that there are SO users active for years that previously produced only few answers now posting over 50 in less than a day and cautioned that the amount of AI generated answers could suffocate SO if everyone starts doing it. \\
\textbf{Evolving AI regulations:} OpenAI TOS policies on their website \cite{Termsofu98:online} says the content co-authored with the OpenAI API policy, creators who wish to publish their first-party written content (e.g., a book, compendium of short stories) created in part with the OpenAI API are permitted to do so under the condition that the
published content must be attributed to the author's name or company, with a clear disclosure of the AI's role in generating the content, ensuring that readers can easily understand the involvement of AI. For instance, according to this statement, one must detail in a Foreword or Introduction (or some place similar) the relative roles of drafting, editing, etc. People should not represent API-generated content as being wholly generated by a human or wholly generated by an AI. It is a human who must take ultimate responsibility for the content being published \cite{christodorescu2024securing}\footnote{ Policies emerging in European Union, the People’s Republic of China, and the United States, as well as the governance efforts in multilateral settings (e.g., G7) are trying to design safeguards into the processes and procedures around using GenAI}. 
\section{Proposing: Generative AI Transparency \& Accountability Evaluation (GATE) Checklist }
 Using a checklist to document any process is not a new concept \cite{minhas2023checklists}. For example, since the 1930s, checklists have been a standard operating procedure for pilots and other aviators in the aviation industry. In medicine, checklists are used as a decision
aid to identify a medical condition and decide on an appropriate course of treatment. In comparison, surgical checklists are recommended as a safety measure to reduce the margin of human error and any adverse effects during surgery \cite{minhas2023checklists}.
In SE, Wieringa et al. \cite{wieringa2012towards} developed a checklist as a guide to performing empirical research effectively. Belli et al. \cite{belli1996towards} developed a checklist to streamline code reviewing. Recently, Patel et al. \cite{patel2024state} proposed a comprehensive release-readiness checklist for GenAI-based Software Products. Which was designed to guide practitioners in evaluating release readiness aspects
such as performance, monitoring, and deployment strategies, aiming to enhance
the reliability and effectiveness of LLM-based applications in real-world settings. Taking inspiration from these works, in this study, we propose a two pronged checklist to guide researchers when using 
 cloud-based LLM tools in various academic tasks that need understanding regarding data protection and awareness regarding legal implications while using GenAI generated content. Table \ref{tableChecklist} is the checklist we propose. Refer to the additional notes in the last column of the table for more details.
\begin{table*}[!htpb]
\vspace{-2mm}
\label{tableChecklist}
\centering
\caption{Generative AI Accountability \&  Transparency Evaluation (GATE) Checklist for risk assessment and research disclosure standards }
\begin{tabular}{p{2cm}p{3cm}p{1cm}p{6cm}}

\multicolumn{4}{c}{\textbf{ Transparency assessment}}  \\ \hline
\multicolumn{1}{l}{Data   legality}            & \multicolumn{1}{l}{Is the   data source legally compliant?}           & \multicolumn{1}{l}{{[} {]} Yes {[}   {]} No} & \begin{tabular}[p{6cm}]{@{}|p{6cm}@{}}- Verify   the dataset’s source (publicly available, licensed, proprietary).\\    - Confirm compliance with data usage rights.\\    - Ensure no copyrighted or sensitive data is used.\\    - Check compliance with regulations (e.g., GDPR, HIPAA).\end{tabular} \\ \hline
\multicolumn{1}{l}{Output   ownership}         & \multicolumn{1}{l}{Is output   ownership clear?}                      & \multicolumn{1}{l}{{[} {]} Yes {[}   {]} No} & \begin{tabular}[p{6cm}]{@{}|p{6cm}@{}}- Check   the GenAI tool's terms of service.\\    - Evaluate output's licensing implications.\\    - Ensure proper attribution for outputs to avoid IP claims.\end{tabular}                                                                           \\ \hline
\multicolumn{1}{l}{Regulatory   compliance}    & \multicolumn{1}{l}{Does the   research comply with AI regulations?}   & \multicolumn{1}{l}{{[} {]} Yes {[}   {]} No} & \begin{tabular}[p{6cm}]{@{}|p{6cm}@{}}- Assess   compliance with local AI regulations (e.g., EU AI Act).\\    - Ensure adherence to ethical AI guidelines.\\    - Check institutional review board (IRB) requirements.\end{tabular}                                                                     \\ \hline
\multicolumn{1}{l}{Licensing   compatibility}  & \multicolumn{1}{l}{Is the   licensing of GenAI outputs compatible?}   & \multicolumn{1}{l}{{[} {]} Yes {[}   {]} No} & \begin{tabular}[p{6cm}]{@{}|p{6cm}@{}}- Verify   compatibility with open-source licenses.\\ - Avoid combining incompatible licenses (e.g., proprietary and open-source).\end{tabular}  \\ \\
\multicolumn{4}{c}{\textbf{Accountability (disclosure standards) assessment}} \\ \hline
\multicolumn{1}{l}{GenAI   usage declaration}  & \multicolumn{1}{l}{Is GenAI   usage disclosed?}                       & \multicolumn{1}{l}{{[} {]} Yes {[}   {]} No} & \begin{tabular}[p{6cm}]{@{}|p{6cm}@{}}- Clearly   state which parts of the research utilized GenAI.\\    - Provide the name and version of the GenAI tool used.\end{tabular}                                                                                                                            \\ \hline
\multicolumn{1}{l}{Output   attribution}       & \multicolumn{1}{l}{Is GenAI   output clearly attributed?}             & \multicolumn{1}{l}{{[} {]} Yes {[}   {]} No} & \begin{tabular}[p{6cm}]{@{}|p{6cm}@{}}-   Distinguish between human-authored and GenAI-generated content.\\    - Attribute outputs (e.g., “Generated using {[}Tool Name{]} Accessed on {[}Date{]}”).\end{tabular}                                                                                                \\ \hline
\multicolumn{1}{l}{Compliance   statement}     & \multicolumn{1}{l}{Is there   an ethical/legal compliance statement?} & \multicolumn{1}{l}{{[} {]} Yes {[}   {]} No} & \begin{tabular}[p{6cm}]{@{}|p{6cm}@{}}- Include   a compliance statement for ethical guidelines and legal requirements.\\    - Disclose and reference data sources.\end{tabular}                                 \\ \hline
\multicolumn{1}{l}{GenAI   contribution}       & \multicolumn{1}{l}{Is   GenAI’s contribution documented?}             & \multicolumn{1}{l}{{[} {]} Yes {[}   {]} No} & \begin{tabular}[p{6cm}]{@{}|p{6cm}@{}}-   Describe GenAI's specific role in the methodology.\\    - Disclose limitations and threats to validity and how they were addressed.\end{tabular}            \\ \hline
\multicolumn{1}{l}{Authorship}                 & \multicolumn{1}{l}{Are   researchers credited, not GenAI?}            & \multicolumn{1}{l}{{[} {]} Yes {[}   {]} No} & \begin{tabular}[p{6cm}]{@{}|p{6cm}@{}}- Ensure   researchers are credited as primary contributors.\\    - Credit GenAI as a tool or assistant..\end{tabular} 

 \\ \hline
\multicolumn{1}{l}{Open Science} & \multicolumn{1}{p{6cm}}{Is source code made public referencing other code repositories used for development?}           & \multicolumn{1}{l}{{[} {]} Yes {[}   {]} No} & \begin{tabular}[p{6cm}]{@{}|p{6cm}@{}} -  Acknowledge other source repositories such as GitHub and Stackoverflow which were used in the code development.  
\end{tabular}                         
\end{tabular}
\end{table*}
\vspace{-3mm}
\section{Related work}
\textbf{GenAI in SE}:
Ebert et al. \cite{ebert2023generative} explored the utility of GenAI for improving software development and software
productivity through code generation, test case
generation from requirements, re-establishing traceability,
explaining code, refactoring of
legacy code, software maintenance with augmented guidance,
Moreover, improving existing code.  However, Ebert et al. caution that while generative AI can help in all these tasks, several risks need to be considered and mitigated.  For example, AI tools can hallucinate, causing privacy and security implications as the code shared with the tool is not open-sourced or, worse, if proprietary, might be used for training leading to grave consequences.  Sauvola et al. \cite{sauvola2024future} analyzed the potential of generative AI and LLM technologies for future software development paths and highlighted the need for new tools to understand the potential, limitations, and risks of generative AI, as well as guidelines for using it.  Carteton et al. \cite{carleton2024generative} highlighted that GenAI brings new ethical dilemmas and intellectual property (IP) challenges.  The ownership of AI-generated code remains ambiguous: Who is responsible for the generated end-product?  AI-assisted creation demands legally sound guidelines to ensure accountability.  From a regulatory perspective, the responsibility of automatically generated codes and content bypassing ethical considerations must be addressed.  \\ 
\textbf{Legal dimensions of GenAI usage}
Weis et al. \cite{weisz2023toward} highlight that the GenAI may have been trained on data-protected
by regulations such as the General Data Protection Regulation (GDPR)\footnote{A legal framework that governs the collection and processing of personal data for individuals in the European Union (EU) and the European Economic Area (EEA)}, which prohibits the re-use of data beyond the purposes for which it was collected. LLMs, often called ``stochastic parrots," can reproduce or remix training data, potentially violating copyrights or imposing restrictive licenses on outputs.  For instance, Codex may generate copyrighted code or code under non-commercial Creative Commons licenses.  A lawsuit against GitHub, Microsoft, and OpenAI highlights these concerns \cite{GitHubCo41:online}.
Fransces et al. \cite{franceschelli2022copyright} explained that the GenAI tools use copyrighted works for training and store copies of protected works for training purposes.  Also, the nature of outputs generated by genAI—unlike traditional AI, which follows explicit rules provided by programmers—relies on techniques and acquired knowledge without direct human intervention.  This complexity makes determining who holds copyright ownership for such outputs challenging.  It was also highlighted that the legal status of using copyrighted works for non-market-encroaching purposes (such as research) remains unclear and may depend on specific circumstances.  Researchers have given special guidelines to clarify their position through TOS if other users could use their generative model, and to keep updated about the evolution of the legislative frameworks at the national and international level \cite{ren2024copyright}. 
\vspace{-3mm}
\section{Conclusion}
This paper investigates the need for a checklist while using cloud-based free-tier GenAI that entails 1) Transparency aspects such as data legality, ethical considerations, licensing and regulatory compliance and 2) Accountability (GenAI generated content usage) related aspects such as authorship and GenAI usage declaration in software engineering research. In the advent of such growing concerns, the proposed checklist can guide researchers in evaluating legal and ethical implications of using GenAI products in research. Through this work, we envision to spark discussion and awareness about the legal risks of GenAI in SE research. Offer a forward-looking vision and actionable steps for researchers to address these challenges and propose a checklist that could serve as a foundation for future work in this emerging area.  






%

\bibliographystyle{ieeetr} 
\bibliography{references}

\end{document}